\documentclass[preprint,12pt]{elsarticle}

\usepackage{amssymb}
\usepackage{amsmath}
\usepackage[linesnumbered,ruled]{algorithm2e}
\usepackage{multirow}
\usepackage{booktabs}
\usepackage{url}
\usepackage{makecell}
\usepackage{hyperref}
\usepackage{geometry}

\journal{Signal Processing}

\begin{document}
\newgeometry{left=3cm, right=3cm, top=3cm, bottom=3cm}
\thispagestyle{empty}
\onecolumn
\null
\vspace*{\fill}
{\large\noindent This is the authors' accepted manuscript. It has been accepted for publication in \textit{Signal Processing} (Elsevier). \href{https://doi.org/10.1016/j.sigpro.2025.110456}{DOI: 10.1016/j.sigpro.2025.110456}

\vspace{\baselineskip}

\noindent\textbf{Copyright:} © 2025. This manuscript version is made available under the CC-BY-NC-ND 4.0 license https://creativecommons.org/licenses/by-nc-nd/4.0/}
\vspace*{\fill}
\vspace*{\fill}
\vspace*{\fill}

\clearpage
\restoregeometry

\begin{frontmatter}

\title{Activity-dependent resolution adjustment for radar-based human activity recognition} 

\author[label1]{Do-Hyun Park} 
\author[label1]{Min-Wook Jeon} 
\author[label1]{Hyoung-Nam Kim\corref{cor1}} 

\affiliation[label1]{organization={School of Electrical and Electronics Engineering, Pusan National University},
            city={Busan},
            postcode={46241}, 
            country={Republic of Korea}}
\cortext[cor1]{Corresponding author}
\begin{abstract}
The rising demand for detecting hazardous situations has led to increased interest in radar-based human activity recognition (HAR). Conventional radar-based HAR methods predominantly rely on micro-Doppler spectrograms for recognition tasks. However, conventional spectrograms employ a fixed resolution regardless of the varying characteristics of human activities, leading to limited representation of micro-Doppler signatures. To address this limitation, we propose a time-frequency domain representation method that adaptively adjusts the resolution based on activity characteristics. This approach adaptively adjusts the spectrogram resolution in a nonlinear manner, emphasizing frequency ranges that vary with activity intensity and are critical to capturing micro-Doppler signatures. We validate the proposed method by training deep learning-based HAR models on datasets generated using our adaptive representation. Experimental results demonstrate that models trained with our method achieve superior recognition accuracy compared to those trained with conventional methods.
\end{abstract}




\end{frontmatter}


\section{Introduction}
\label{Sec1}
Recently, radar systems have gained significant attention due to their wide applicability across various fields, including elderly care and healthcare. Consequently, radar-based human activity recognition (HAR) systems have emerged as privacy-preserving technologies capable of promptly detecting dangerous situations, such as falls, thereby enhancing safety and quality of life \cite{ref-falldet2}.
Radar-based HAR systems recognize human activities by analyzing the micro-Doppler effect embedded in target-echo signals. The micro-Doppler effect represents the frequency shifts caused by the minute movements of a target, and it serves as a key feature for identifying various human activities. 

HAR systems employ signal processing techniques that appropriately represent received signals to effectively analyze the movements of different body parts. Signals are primarily transformed into two-dimensional data using various preprocessing methods. For instance, the received signal can be represented in the time--frequency domain (also known as Doppler--time domain, commonly referred to as a spectrogram) \cite{ref-SIGPRO}, range--time domain \cite{ref-RTHAR}, or range--Doppler domain \cite{ref-RDHAR}. Despite these diverse approaches, numerous studies utilize spectrograms for HAR systems because they effectively capture the movement characteristics of different body parts  \cite{ref-falldet3}.

The accuracy of spectrogram-based HAR systems can be strongly influenced by the time--frequency domain representation methods. Current time--frequency analysis methods employed in HAR systems can be broadly classified into two categories: linear and quadratic time--frequency analyses. Linear time--frequency analysis typically involves the short-time Fourier transform (STFT), in which the time-domain signal is segmented into fixed-length windows, and each segment is transformed via a Fourier transform to yield time--frequency domain information. However, the fixed-length window function used in the STFT results in a trade-off between the time and frequency resolution.
Quadratic time--frequency analysis includes methods such as the Wigner--Ville distribution (WVD), smoothed pseudo WVD (SPWVD), and reduced interference distribution using the Hanning kernel (RIDHK). Although the WVD provides a high time--frequency resolution, it experiences cross-term interference problems. SPWVD is an advanced method derived from WVD, and it effectively suppresses cross-term interference while maintaining a high time--frequency resolution, thereby allowing for accurate signal feature representation. The RIDHK provides a higher time--frequency resolution compared with the SPWVD but with increased cross-term interference. As each of these techniques provides diverse advantages, ongoing research has explored the use of the SPWVD and RIDHK in HAR applications \cite{ref-SRHAR1, ref-SRHAR2, ref-SRHAR3}.

The aforementioned time--frequency representation methods are useful for visualizing various micro-Doppler signatures. However, they have limitations in analyzing subtle patterns associated with human activities. Existing time--frequency analysis methods, which use fixed window sizes, exhibit a linearly distributed resolution for time and frequency. This makes it challenging to capture detailed features from the micro-Doppler effects caused by fine movements such as the vibration or rotation of body parts. 

To address these limitations, this study proposes a time--frequency representation technique that is optimized for analyzing distinctive micro-Doppler signature patterns associated with human activities.
The proposed method adaptively adjusts the resolution while maintaining a fixed spectrogram size by: (i) automatically identifying critical frequency ranges corresponding to significant micro-Doppler signatures, (ii) increasing the resolution within these identified ranges, and (iii) decreasing the resolution in less important regions. We validate the superiority of the proposed method by comparing the performance of deep learning models trained on datasets generated using conventional time–frequency techniques and the proposed technique. This comparison demonstrates the effectiveness of the proposed method in improving the HAR accuracy.

The rest of this paper is organized as follows: Section \ref{Sec2} describes the preprocessing technique for frequency-modulated continuous-wave (FMCW) radar and the method for generating spectrograms. Section \ref{Sec3} describes how the resolution of the spectrogram is adaptively adjusted based on activity. In Section \ref{Sec4}, we evaluate the performance of the proposed method using real-world data and compare it against existing methods. Finally, Section \ref{Sec5} concludes the paper.

\section{Preprocessing for FMCW Radar-based HAR}
\label{Sec2}
FMCW radar-based HAR systems perform a preprocessing step to extract features related to fine human movements from raw radar signals. First, the received raw data are arranged into a two-dimensional data matrix. Here, one axis represents the fast time, which corresponds to the time per sweep, and the other axis represents the slow time, which indicates the index of chirps over the measurement duration. Then, a Fourier transform is applied to the fast time domain to yield \(\textbf{x}(r,n)\), which contains the magnitude of the \(r\)-th range bin at the \(n\)-th chirp. The distance information proportional to the delay time of the target echo can be extracted from \(\textbf{x}(r,n)\) \cite{ref-Preproc2}. A fourth-order high-pass Butterworth filter with a cutoff frequency of 0.01 Hz is applied to \(\textbf{x}(r,n)\) to remove static clutter.

\(\textbf{x}(r,n)\) represents the range--slow-time signal obtained after the fast-time Fourier transform and clutter removal. Since \(\textbf{x}(r,n)\) contains only slow-time information, direct frequency analysis would discard temporal variations. Therefore, time–frequency analysis is applied to preserve both temporal and spectral characteristics that are essential for representing micro-Doppler signatures. Time--frequency analysis is performed in the slow time domain of filtered \(\textbf{x}(r,n)\) to generate a spectrogram composed of time and frequency domains, thereby allowing for the visualization of micro-Doppler signatures. The spectrogram, \(\textup{SPEC}(t,f)\), can be obtained using the STFT, as follows:
\begin{equation}
\textup{SPEC}(t,f)=\left| \sum_{n}\sum_{r=r_{s}}^{r_{e}}\textbf{x}(r,n)w(n-t)e^{-j 2 \pi f n} \right|^{2},
\label{eq1}
\end{equation}
where \(r_{s}\) and \(r_{e}\) represent the start and end range bins for the analysis, respectively, and \(w(\cdot)\) denotes the window function. Using a spectrogram removes absolute phase and applies time--frequency smoothing caused by windowing and grid sampling. These effects do not degrade the micro-Doppler information but instead yield a stable time--frequency energy pattern.

\begin{figure}[]
\centering
\includegraphics[width= \textwidth]{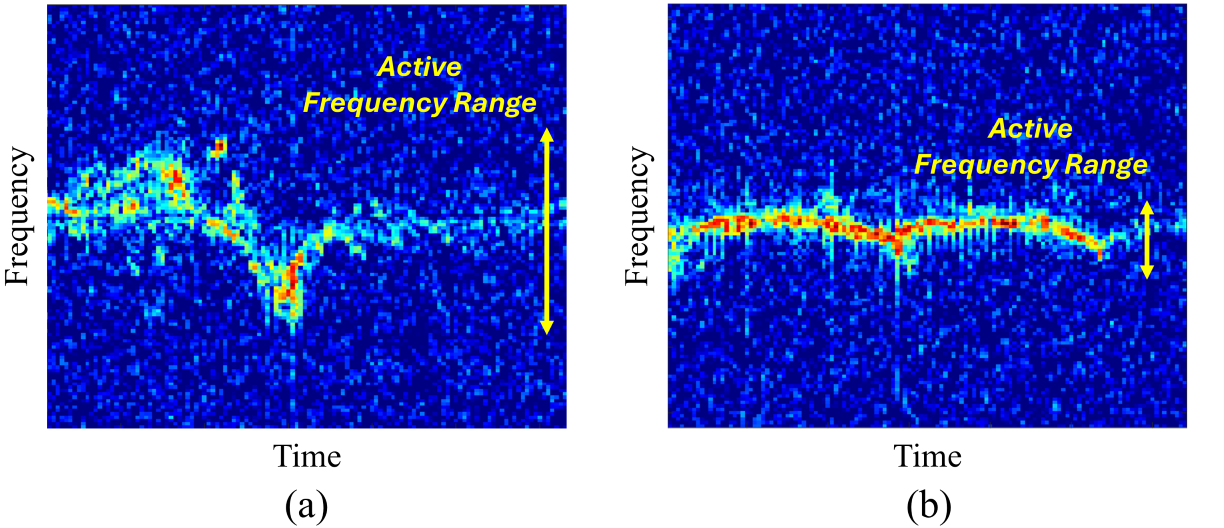}
\caption{Examples of spectrograms showing distinct micro-Doppler signatures obtained from various human activities: (a) falling, (b) limping.}
\label{fig1}
\end{figure}

Fig. \ref{fig1} presents the examples of spectrograms generated from signals containing micro-Doppler effects caused by various human activities. Different human movements generate unique micro-Doppler signatures, which are clearly visible in the spectrograms. However, the fixed time and frequency resolution of the STFT poses a limitation in optimizing the analysis of micro-Doppler signatures associated with diverse human activities. This limitation is due to the varying frequency ranges generated by different actions. For instance, as shown in Fig. \ref{fig1}(a), a falling action results in a micro-Doppler signature distributed across a wide frequency range, making it relatively straightforward to observe. In contrast, the limping movement shown in Fig. \ref{fig1}(b) exhibits a narrower frequency range, making the features within its micro-Doppler signature less sharply defined. Although Fig. \ref{fig1}(b) contains more high-amplitude pixels within the active range, the spectrogram provides coarser frequency resolution, which causes nearby micro-Doppler components to broaden and partially merge. Consequently, the fine time--frequency patterns are less well resolved despite the higher overall magnitude.

\begin{algorithm}[t]
    \SetKwInOut{Input}{Input}
    \SetKwInOut{Output}{Output}
    \Input{\(\text{SPEC}(t, f)\), \(M\), \(f_\mathrm{max}\)}
    \Output{\(\text{SPEC}_{\mathrm{RA}}(t, m)\)}
    
    \(e(f)\leftarrow\sum_{t}\log(\text{SPEC}(t,f))\)
    
    \(f_{nc},f_{pc} \leftarrow \min_{f_{1},f_{2}} J(f_{1},f_{2})\)

    \(f_{c} \leftarrow \max\left(|f_{nc}|,f_{pc}\right)\)

    Calculate \(S(f_\mathrm{max})\) using (\ref{eq4})

    Compute \(P_{m}\) for \(m=0,1,\ldots,M+1\) as evenly spaced points between \(0\) and \(S(f_\mathrm{max})\)

    Convert \(P_{m}\) to \(p_{m}\) using (\ref{eq5})

    \For{{iteration \(m=1,\ldots,M\)} \do}
      {
          \For{{iteration \(f=0,\ldots,f_\mathrm{max}\)} \do}
      {
        Compute \(m\)-th filter \(\text{FB}(m,f)\) using (\ref{eq6})
      }
}

    \(\text{SPEC}_\mathrm{pos}(t,m) \leftarrow \sum_{f=0}^{f_\mathrm{max}}\text{FB}(m,f) \cdot \text{SPEC}(t,f) \)
    
    \(\text{SPEC}_\mathrm{neg}(t,m) \leftarrow \sum_{f=0}^{f_\mathrm{max}}\text{FB}(m,f) \cdot \text{SPEC}(t,-f) \)

Construct \(\text{SPEC}_{\mathrm{RA}}(t, m)\) by combining \(\text{SPEC}_\mathrm{pos}(t, m)\) and \(\text{SPEC}_\mathrm{neg}(t, m)\)

    \caption{Resolution-adjusted spectrogram generation}
    \label{algorithm}
\end{algorithm}

\section{Activity-dependent Resolution Adjustment}
\label{Sec3}
This study proposes a novel, activity-dependent adaptive time--frequency analysis method to enhance the accuracy of micro-Doppler signature recognition. The proposed technique consists of two steps: (i) identifying the frequency range of micro-Doppler signatures; and (ii) nonlinearly adjusting the frequency-domain resolution of the spectrogram within the identified frequency range. Specifically, the method increases resolution in critical frequency ranges and decreases it in less important regions, while maintaining a fixed spectrogram size. This allows for feature analysis customized to the characteristics of micro-Doppler signatures varying with each activity.

Algorithm \ref{algorithm} describes how the conventional spectrogram is transformed into a resolution-adjusted (RA) spectrogram, $\text{SPEC}_\mathrm{RA}(t,m)$. The inputs for this algorithm are the spectrogram, \(\text{SPEC}(t,f)\), number of filters in the filter bank, \(M\), and the maximum frequency of the spectrogram, \(f_\mathrm{max}\). In the first step, the spectrogram undergoes a logarithmic transformation and is then projected onto the time axis to obtain the energy distribution, \(e(f)\), as follows:
\begin{equation}
e(f) = \sum_{t} \log(\text{SPEC}(t,f)).
\end{equation}
This helps clearly identify the energy distribution of the micro-Doppler signatures in the frequency domain.

To explore the active frequency range corresponding to different activities, the negative corner frequency, \(f_{nc}\), and positive corner frequency \(f_{pc}\), are determined by solving the optimization problem defined as:
\begin{equation}
\begin{aligned}
(f_{nc}, f_{pc}) = \arg&\min_{f_1, f_2} J(f_1, f_2),\\ \text{subject to}\quad -f_{\text{max}}& < f_1 < f_2 < f_{\text{max}},
\end{aligned}
\end{equation}
where the objective function \( J(f_1, f_2) \) is defined as the sum of logarithmic mean square (LogMS) energies calculated over three distinct intervals: \([ -f_{\text{max}}, f_1 ]\), \([ f_1, f_2 ]\), and \([ f_2, f_{\text{max}} ]\). The LogMS over an interval from \( n \) to \( m \) is computed as follows:
\begin{equation}
\textrm{LogMS}(n,m) = (m - n + 1)\log\left( \frac{1}{m - n + 1}\sum_{i = n}^{m} e^{2}(i)\right).
\end{equation}
Minimizing this function concentrates micro-Doppler signature energy within the interval [\(f_1\), \(f_2\)], clearly separating it from noise and irrelevant frequency components. To maintain the stability of the resolution adjustment, the corner frequency, \(f_{c}\), is selected as the larger value between \(|f_{nc}|\) and \(f_{pc}\).

As a part of the nonlinear resolution adjustment, the linear frequency points from the spectrogram are mapped to their corresponding positions on the nonlinear frequency scale. The scaling function that transforms the linear frequency component, \(f\), into the nonlinear frequency scale, \(S(f)\), is defined as follows:
\begin{equation}
S(f) = \dfrac{f_{c}}{\log(2)} \log\left( 1+ \dfrac{f}{f_{c}} \right).
\label{eq4}
\end{equation}
This nonlinear mapping was chosen based on the empirical observation that micro-Doppler signatures of human activities predominantly concentrate in lower-frequency regions. The scaling function provides a higher resolution in the frequency range below \(f_c\) by mapping it onto a denser, nearly linear scale. In contrast, the frequency range above \(f_c\) is sparsely analyzed at a nonlinear scale, resulting in a lower resolution.

After \(S(f_\mathrm{max})\) is computed with the scaling function, nonlinear frequency points, \(P_m\), are generated with \(M+2\) uniformly distributed points between 0 and \(S(f_\mathrm{max})\). Points \(P_m\) are mapped back to their corresponding linear frequency points, \(p_m\), by applying the following inverse transformation:

\begin{equation}
p_{m} = f_{c}\cdot\left( 10^{z} -1 \right), \quad
z = \frac{\log(2) \cdot P_{m}}{f_{c}}.
\label{eq5}
\end{equation}
This inverse mapping is chosen to reconstruct the linear frequency domain while maintaining the desired nonuniform resolution of the nonlinear scale. It preserves a fixed input size and emphasizes the lower-frequency region where micro-Doppler energy is typically concentrated, thereby improving representational efficiency without additional computational cost.

To facilitate resolution adjustment using \(p_{m}\), the filter bank, \(\text{FB}(m,f)\), is derived as follows:
\begin{equation}
\text{FB}(m,f) = 
\begin{cases} 
\frac{f - p_{m-1}}{p_{m} - p_{m-1}}, & \text{if } p_{m-1} \leq f \leq p_{m} \\[0.5em]
\frac{p_{m+1} - f}{p_{m+1} - p_{m}}, & \text{if } p_{m} < f \leq p_{m+1} \\[0.5em]
0, & \text{otherwise}
\end{cases}
\label{eq6}
\end{equation}
The filters in the filter bank are designed such that for each frequency point, they linearly increase from 0 to 1 between the previous and current frequency points and linearly decrease from 1 to 0 between the current and subsequent frequency points. Triangular filters are adopted for their locality and efficiency and for their ability to yield an approximately constant total gain when overlapped, which helps avoid unnecessary spectral distortion. This filter bank is independently applied to the positive and negative frequency components of the spectrogram. The processed components are then recombined to generate the RA spectrogram.

\begin{figure}[!t]
\centering
\includegraphics[width= \textwidth]{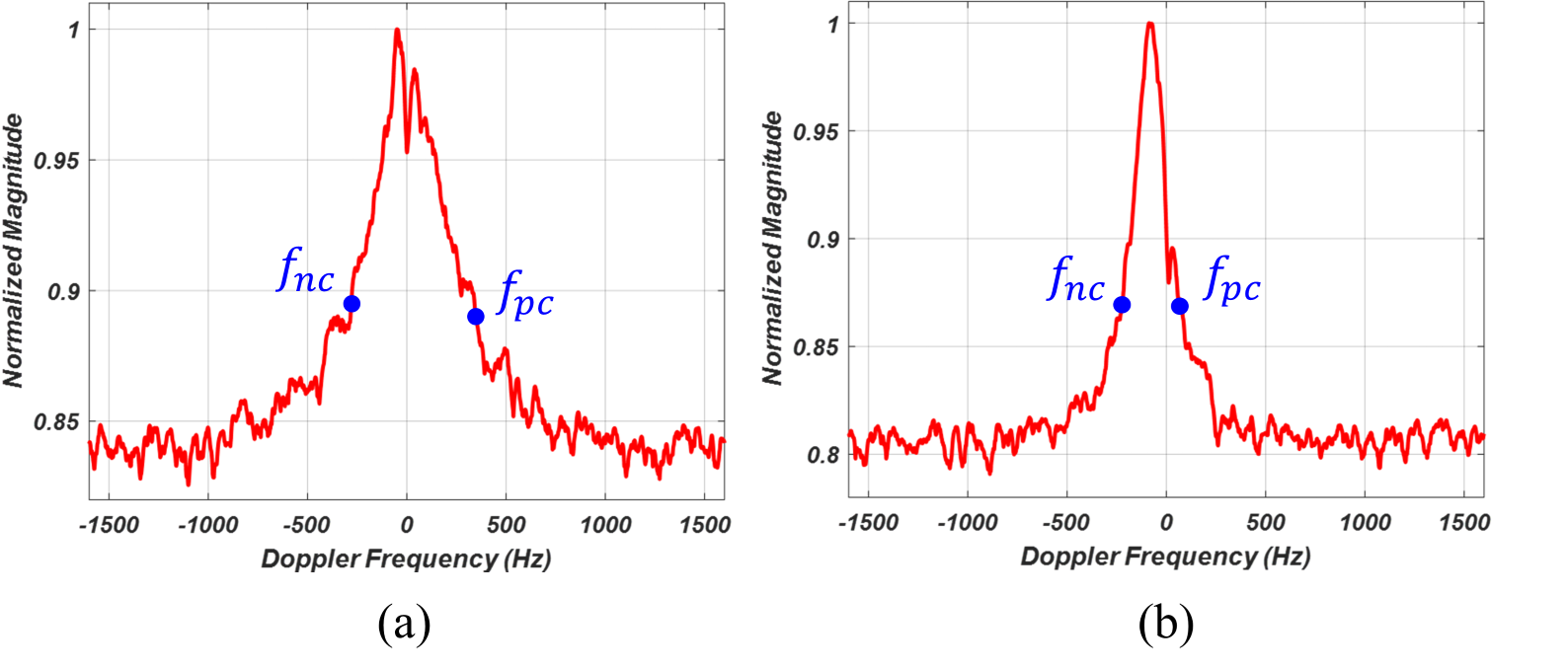}
\caption{Energy distributions and corner frequency estimation results for various activities: (a) falling, (b) limping.}
\label{fig2}
\end{figure}

\begin{figure}[!t]
\centering
\includegraphics[width= \textwidth]{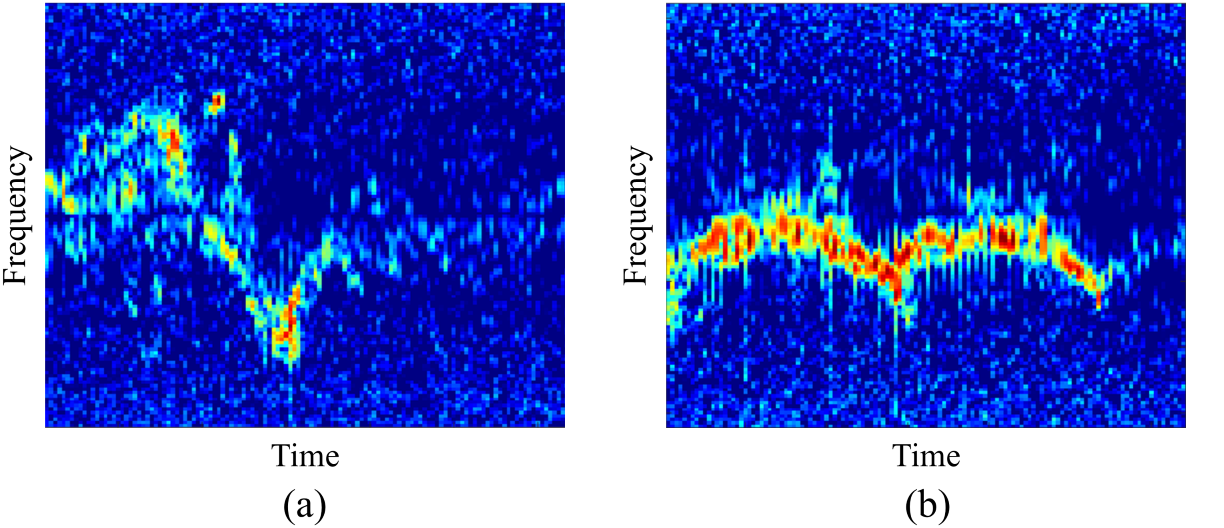}
\caption{RA spectrograms generated from various human activities: (a) falling, (b) limping.}
\label{fig3}
\end{figure}

\begin{figure}[!t]
\includegraphics[width= \textwidth]{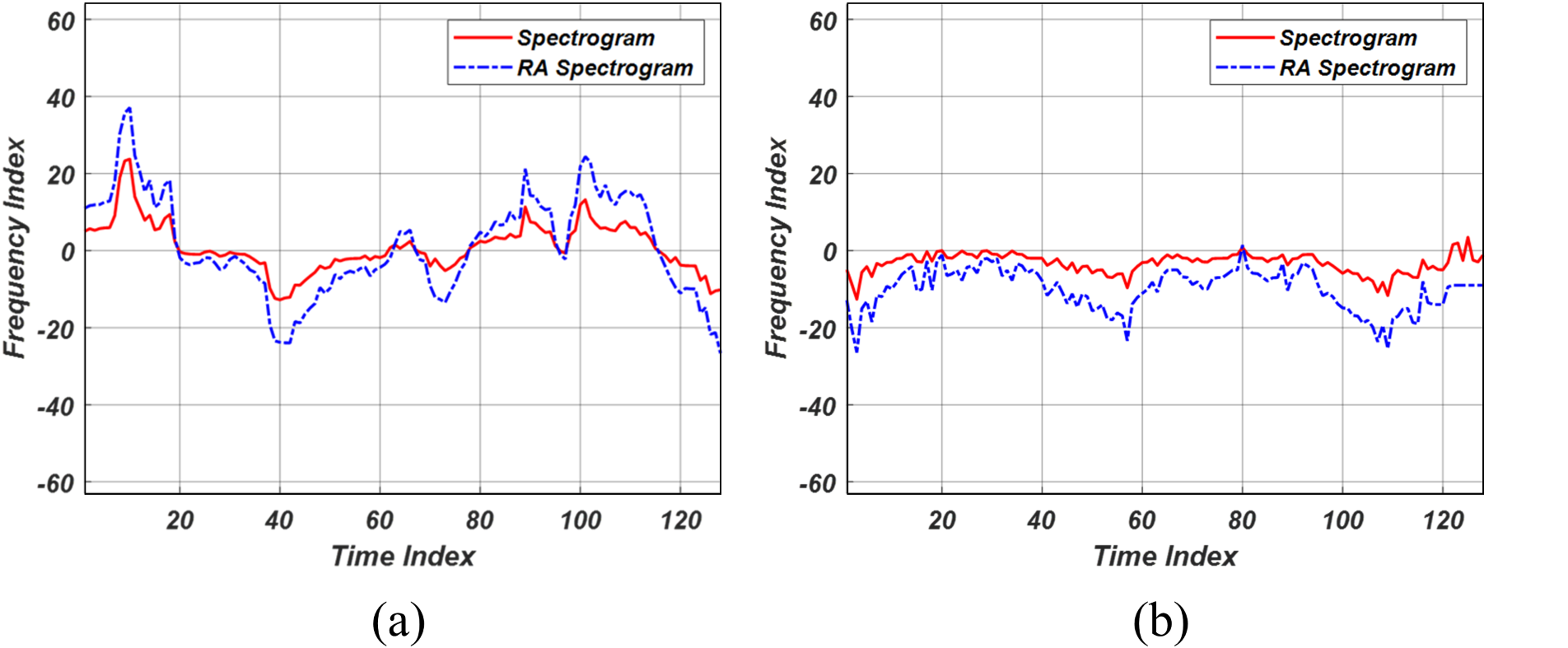}
\centering
\caption{Micro-Doppler signature tracking results for various activities: (a) falling, (b) limping.}
\label{figA}
\end{figure}

Figs. \ref{fig2} and \ref{fig3} illustrate the process and outcomes of generating RA spectrograms for various activities using the proposed method, based on the same dataset as Fig. \ref{fig1}. Specifically, Fig. \ref{fig2} highlights the identification of negative and positive corner frequencies associated with the falling and limping motions. The corner frequency is subsequently employed to adjust the resolution. The examples of the RA spectrogram shown in Fig. \ref{fig3} illustrate that the enhanced resolution within the frequency range corresponding to micro-Doppler signatures allows for precise pattern analysis. 

Fig. \ref{figA} presents the tracking results of micro-Doppler signatures in the conventional and RA spectrograms, which highlight the extent of resolution improvement. Signature tracking is performed by extracting the frequency index with the highest value for each time index within each spectrogram, followed by the application of a Kalman filter. This comparison demonstrates that the adaptive resolution approach effectively reveals subtle micro-Doppler features previously indistinct under fixed resolutions, enhancing clarity and distinguishability. It should be noted that in Figs. \ref{fig3}(b) and \ref{figA}(b), the improved resolution substantially enhances the clarity of signature features for activities such as limping, which are primarily concentrated in the low-frequency band. These findings demonstrate the strong impact of resolution enhancement in such cases.

\section{Performance Analysis}
\label{Sec4}
We collected data for various daily activities to evaluate the performance of the proposed method. The data were acquired using the Texas Instruments AWR1642BOOST FMCW radar. Participants performed six different daily activities (A1 to A6: falling, limping, picking up objects, running, sitting, and walking) within a range of 4 m with respect to the radar. To ensure the diversity of the dataset, five participants (P1 to P5) contributed to the data collection process. The dataset contains 2,250 samples with the following class distribution: fall 256, limping 333, picking 456, running 330, sitting 311, and walking 564.

Given the small dataset size, we employed two evaluation methods. (i) Leave-one-out cross-validation (LOOCV): The data obtained from four participants were utilized to train the deep learning-based recognition model. The data obtained from one participant, which were not included in the training dataset, were used for testing. This approach ensured an unbiased assessment of the generalization capability of the model. (ii) Stratified 5-fold cross-validation: The dataset was divided into 5 folds while preserving the class proportions to address potential bias due to class imbalance. In each iteration, 4 folds were used for training and one for testing, and the process was repeated 5 times so that every sample was tested exactly once. The final performance was reported as the mean accuracy across all folds, providing a stable and reliable estimate of model performance under balanced data conditions.

We compared the recognition performances of deep learning-based HAR models trained on datasets obtained using conventional time--frequency analysis methods with those trained on datasets obtained using the proposed method. It should be noted that the number of filters for RA spectrogram generation was set to ensure identical dimensions between conventional and proposed methods, thereby maintaining equal computational complexity. The input image data were log-transformed and preprocessed by normalizing them to have a mean of 0 and variance of 1. For the HAR models, we utilized convolutional neural network (CNN) models that were previously used in HAR research \cite{ref-CNN, ref-ConvNet}. In addition, we used CNN models that are widely used in image classification, such as ResNet18 \cite{ref-RESNET}, VGG16, and VGG19 \cite{ref-VGG}.

\begin{figure}[]
\centering
\includegraphics[width= \textwidth]{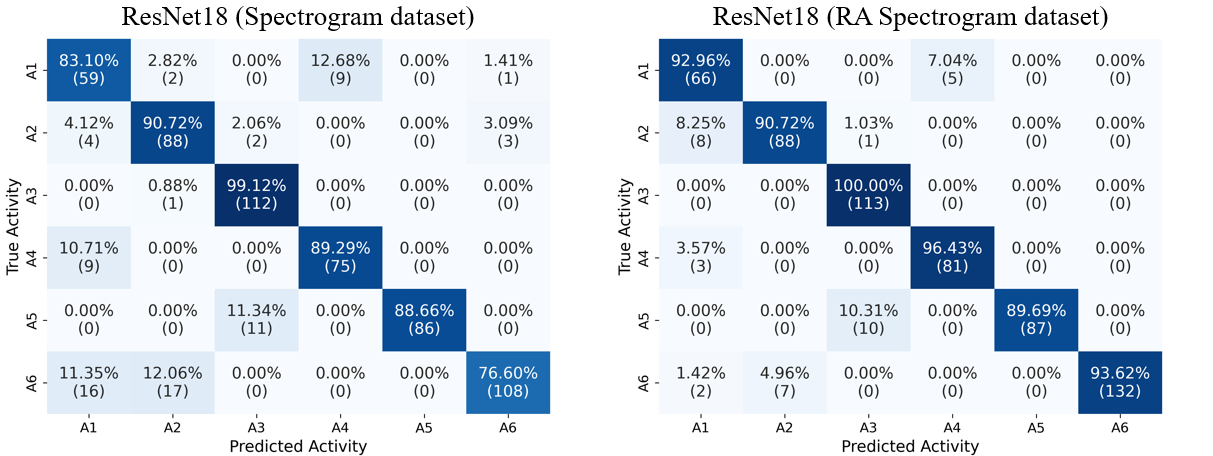}
\caption{Confusion matrices for model trained on conventional and RA spectrogram datasets.}
\label{fig_conf}
\end{figure}

Fig. \ref{fig_conf} presents the confusion matrices that compare actual activities with those predicted by the HAR model. The ResNet18 model was trained on datasets generated using conventional and proposed RA spectrograms. Training datasets included data from participants P2–P5, while the test dataset comprised data from participant P1. Results indicate that the model trained on RA spectrograms achieves higher average recognition accuracy than the one trained on conventional spectrograms. In particular, recognition accuracy significantly improved for activity A1 (falling), a critical factor in HAR due to its importance in safety and health monitoring applications.

\begin{table}[]
\centering
\includegraphics[width= 0.9 \textwidth]{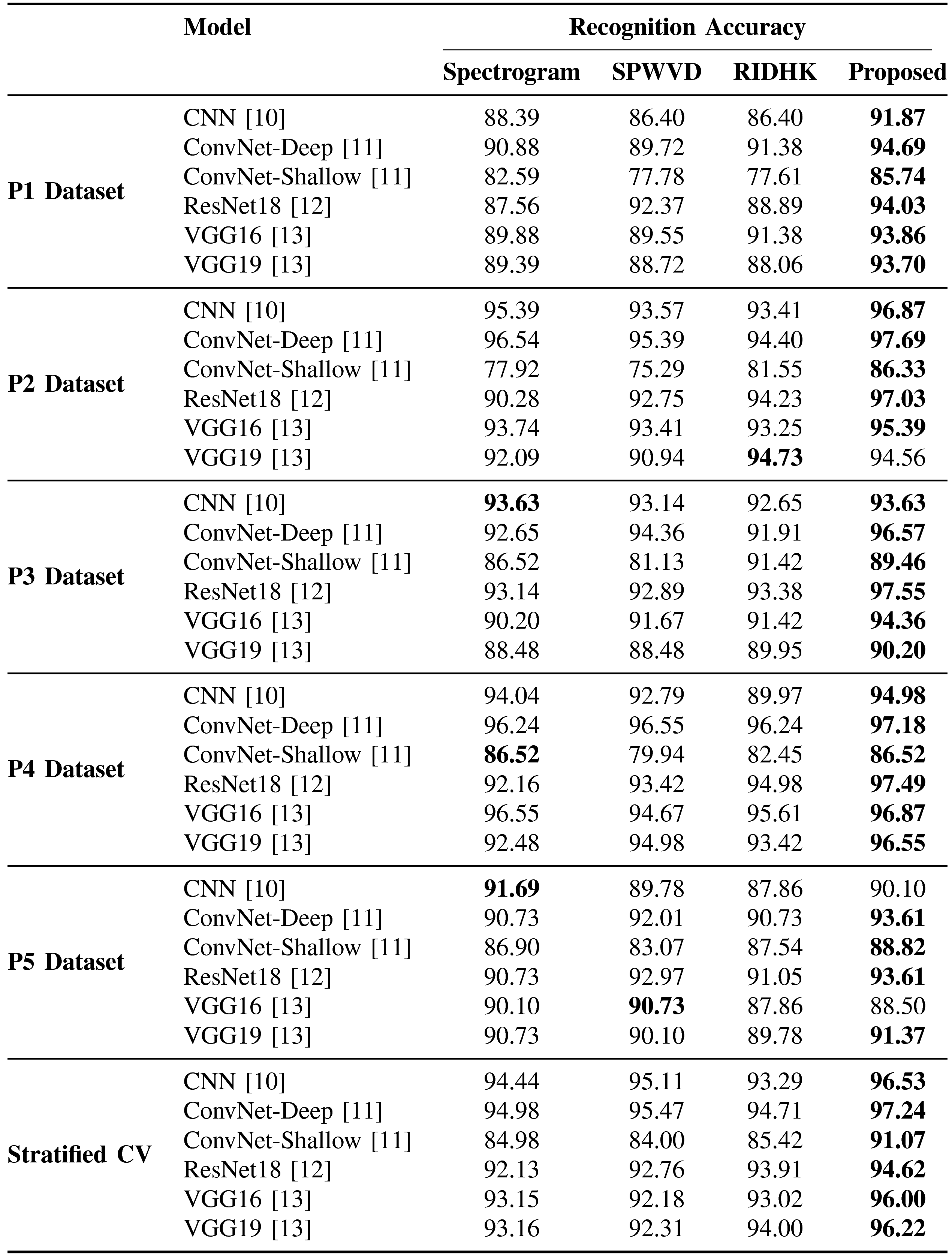}
\caption{Accuracy of HAR models for various datasets}
\label{table1}
\end{table}

Table \ref{table1} presents the average recognition accuracies for the models based on various time–frequency analysis methods. The proposed RA spectrogram consistently outperforms the conventional spectrogram, SPWVD, and RIDHK representations across all datasets (P1--P5) and under the stratified 5-fold cross-validation (CV). Based on the LOOCV results over the P1--P5 datasets, the proposed method improves the recognition accuracy by 0.9 \%, 2.5 \%, 3.3 \%, 5.2 \%, 1.7 \%, and 2.6 \% for CNN, ConvNet-Deep, ConvNet-Shallow, ResNet18, VGG16, and VGG19, respectively, compared with conventional spectrograms. Under the stratified CV dataset, the proposed approach yields an average improvement of 3.1 \%, 3.3 \%, and 2.9 \% over the conventional spectrogram, SPWVD, and RIDHK representations, respectively, when averaged across all tested models. These results demonstrate that the proposed method provides a refined resolution suitable for human activity analysis, enabling more effective capturing and classification of micro-Doppler signature patterns compared with existing methods.

\section{Conclusion}
\label{Sec5}
We proposed an adaptive time–frequency representation method that dynamically adjusts its resolution based on activity-induced variations in micro-Doppler signatures. The proposed method identifies the relevant micro-Doppler frequency regions and selectively enhances their resolution while reducing resolution elsewhere, maintaining a fixed spectrogram size to effectively represent signature characteristics. Experimental results demonstrate that the proposed approach achieves a higher accuracy compared with various time--frequency analysis techniques. Future research will focus on extending the proposed resolution adjustment method beyond the time–frequency domain to other signal representation domains.

\section{Funding}
This work was supported by the National Research Foundation of Korea(NRF) grant funded by the Korea government(MSIT) (RS-2025-00557790).

\section{Availability of data and materials}
Source code and sample datasets are available in the GitHub repository: \\github.com/Signal-Park/Activity-Dependent-Spectrogram

\end{document}